\documentclass[conference]{IEEEtran}
\usepackage{blindtext}
\usepackage{graphicx}

\usepackage[labelfont=bf]{caption}
\usepackage{pgfplots}
\usepackage{pgfplotstable}
\pgfplotsset{compat=1.15}
\usepackage{filecontents}
\usepackage{tikz}
\usetikzlibrary{patterns}
\usetikzlibrary{intersections}
\usetikzlibrary{calc}
\usepackage{color}
\usepackage[strings]{underscore}
\usepackage{subcaption}

\usepackage[linesnumbered,ruled]{algorithm2e}
\usepackage{algpseudocode}
\usepackage{url}
\usepackage{footnote}
\makesavenoteenv{tabular}
\makesavenoteenv{table}
\usepackage{amsmath,amsthm}

\usepackage{booktabs}
\usepackage{multirow}
\usepackage{comment}
\usepackage{pifont}
\usepackage{pgf-pie}
\usetikzlibrary{shadows}

\usepackage{enumitem}
\usepackage{siunitx}
\usepackage{listings}
\usepackage{parcolumns}
\usepackage{multicol}
\usepackage[utf8]{inputenc}
\usepackage{array}
\DeclareUnicodeCharacter{2212}{-}
\usepackage[english]{babel}
\usepackage{colortbl}
\usepackage{tcolorbox}  
\usepgfplotslibrary{groupplots}
\usepackage{xspace}
\usepackage{tabularx}
\usepackage{xpatch}

\usepackage{simplebnf}

\xpatchbibmacro{name:andothers}{%
  \bibstring{andothers}%
}{%
  \bibstring[\emph]{andothers}%
}{}{}

\usepackage{xspace}
\newcommand{\MyPara}[1]{\vspace{.3em}\noindent\textbf{\textit{#1}}~~}
\newcommand{\codefont}[1]{{\texttt{#1}}}

\newcommand{\eg}{\emph{e.g.,}\xspace}

\newcommand{\naively}[0]{na\"{i}vely\xspace}

\newcommand{\tool}{{\small{\textsc{GRAphRef}}}\xspace}

\newcommand{\afl}{{\small{\textsc{AFL}}}\xspace}

\usepackage{xcolor}
\usepackage[normalem]{ulem}
\newif\ifshowchanges
\showchangesfalse

\ifshowchanges
    \newcommand{\added}[1]{\textcolor{blue}{#1}}
    \newcommand{\deleted}[1]{\textcolor{red}{\sout{#1}}}
    
    \newcommand{\edit}[1]{\textcolor{blue}{#1}}
    \newcommand{\reviewA}[1]{\textcolor{teal!80!black}{#1}}
    \newcommand{\reviewB}[1]{\textcolor{orange!80!black}{#1}}
    \newcommand{\reviewC}[1]{\textcolor{purple!70!black}{#1}}
\else 
    \newcommand{\added}[1]{#1}
    \newcommand{\deleted}[1]{}
    
    \newcommand{\edit}[1]{#1}
    \newcommand{\reviewA}[1]{#1}
    \newcommand{\reviewB}[1]{#1}
    \newcommand{\reviewC}[1]{#1}
\fi

\usepackage{makecell}
\usepackage{flushend}
\usepackage{balance}

\usepackage{hyperref}
\hypersetup{hidelinks}

\usepackage[T1]{fontenc}
\usepackage[utf8]{inputenc}

\title{Generating Highly Structured Test Inputs Leveraging Constraint-Guided Graph Refinement}

\author{
    \IEEEauthorblockN{Zhaorui Yang\IEEEauthorrefmark{1}, Yuxin Qiu\IEEEauthorrefmark{1}, Haichao Zhu\IEEEauthorrefmark{2}, Qian Zhang\IEEEauthorrefmark{1}}
    \IEEEauthorblockA{\IEEEauthorrefmark{1}
    University of California Riverside, USA\\
    \{zyang247, yqiu054, qzhang\}@ucr.edu}
    \IEEEauthorblockA{\IEEEauthorrefmark{2} Meta, USA\\
    hczhu@reality.vision}
}

\begin{document}

\maketitle

\begin{abstract}
\edit{[Context]} Modern AI applications increasingly process highly structured data, such as 3D meshes and point clouds, where test input generation must preserve both structural and semantic validity. However, existing fuzzing tools and input generators are typically handcrafted for specific input types and often generate invalid inputs that are subsequently discarded, leading to inefficiency and poor generalizability. 
\edit{[Objective]} This study investigates whether test inputs for structured domains can be unified through a graph-based representation, enabling general, reusable mutation strategies while enforcing structural constraints. We will evaluate the effectiveness of this approach in enhancing input validity and semantic preservation across eight AI systems.
\edit{[Method]} We develop and evaluate \tool, a graph-based test input generation framework that supports constraint-based mutation and refinement. \tool maps structured inputs to graphs, applies neighbor-similarity-guided mutations, and uses a constraint-refinement phase to repair invalid inputs. We will conduct a confirmatory study across eight real-world mesh-processing AI systems, comparing \tool with AFL, MeshAttack, Saffron, and two ablated variants. Evaluation metrics include structural validity, semantic preservation (via prediction consistency), and performance overhead. \reviewB{Experimental data is derived from ShapeNetCore mesh seeds and model outputs from systems like MeshCNN and HodgeNet. Statistical analysis and component latency breakdowns will be used to assess each hypothesis.}
\end{abstract}

\flushend

\section{Introduction}\label{sec:introduction}



The rapid evolution of modern software has reshaped numerous domains, such as medical diagnostics~\cite{CancerImage}, autonomous vehicles~\cite{apolloAuto}, and 3D animation~\cite{hanocka2019meshcnn}. 
Reliability in these systems is critical, as their failures can result in severe consequences that include financial loss and threats to human life~\cite{lou2022testingADS}.

However, testing such emerging applications is challenging because it heavily relies on abundant, highly structured, and semantically rich test inputs.
For example, testing an autonomous driving system requires handling diverse input types across its components~\cite{lou2022testingADS}.
The object detection component processes RGB images formatted as 2D pixel arrays~\cite{RGB}, and the 3D segmentation component~\cite{3dseg} only accepts point clouds.

\MyPara{Current Practices and Limitations.}
One common testing practice is to manually craft or collect inputs, which is time-consuming and often fails to cover all execution scenarios.
Therefore, coverage-guided fuzz testing~\cite{afl, saffron} has emerged and become the de facto default technique for testing large software systems.
Typically, it executes the program with seed inputs, randomly mutates existing inputs to improve a given guidance metric such as branch coverage, and repeats this process of input mutation and program execution to explore diverse program behaviors.
However, \naively applying existing tools is still ineffective and inefficient for generating highly structured test inputs.
\begin{itemize}
    \item {\em Structure Preservation.} Real-world systems that process structured inputs often include a format-parsing stage before executing deeper program logic. For example, Open3D~\cite{open3d-paper} expects mesh inputs in the \codefont{obj} format, where each vertex must begin with the prefix \codefont{v} and each face with \codefont{f}. Applying fuzz testing directly to mesh inputs can disrupt this special syntax.
    \item {\em Extensibility.} Grammar- and generator-based fuzzing techniques have been developed to ensure the syntactic validity of structured inputs. However, they are typically confined to one specific data type, as the grammars or templates are carefully tailored for a particular format. Extending them to new input types often requires significant manual effort.
    \item {\em Testing Efficiency.} Existing fuzzing workflows discard inputs that are invalid or fail to improve guidance metrics. However, our experiments showed that 65\% of the time spent using \afl to generate meshes for MeshCNN~\cite{hanocka2019meshcnn} was wasted on rejected meshes, significantly reducing testing efficiency.
\end{itemize}


To enable efficient and effective generation of structured inputs, we have three core hypotheses.
First, many complex inputs share common graph structures, which can be captured using extensible APIs. Therefore, we can unify them into a graph representation, enabling the reuse of a single mutation engine across input types. Second, input semantics can be preserved by applying similar mutations to structurally similar neighbors, guided by graph locality. Third, many constraints, such as format, connectivity, and geometry, can be encoded and enforced during mutation to ensure input validity.

\MyPara{\tool}
To test the above hypotheses, we propose \tool, a fuzz testing framework for generating structured inputs through graph-based mutation and constraint refinement.
It consists of three main components:


{\em 1. Constraint Enforcement.}
\tool includes a constraint analyzer and verifier. Constraints are defined using a DSL over the graph structure, capturing value-level (e.g., pixel values in [0, 255]), connectivity (e.g., 4-way adjacency in image grids), and structural properties (e.g., manifoldness in triangle meshes). During mutation, the verifier detects violations and applies corrective actions to maintain input validity.

\begin{figure}[t]
    \centering
    \scalebox{0.8}{
  \includegraphics[width=\linewidth]{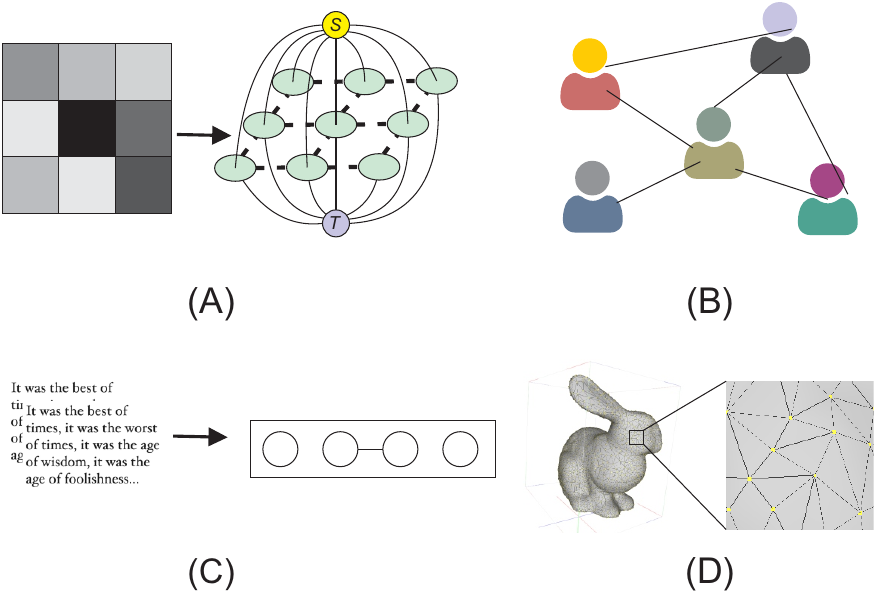}
  }
  \caption{ Graph Representations of Different Data Types: (A) Image as a grid graph where pixels are vertices and edges connect neighbors. (B) Social network with users as vertices and friendships as edges. (C) Text as a sequential graph with characters as vertices and edges for order. (D) Triangle mesh with 3D points as vertices and edges forming the mesh.}
  \label{fig:graph_examples}
\end{figure}

{\em 2. Graph Converter.}
\tool converts structured inputs—such as images, triangle meshes, point clouds, and text—into standardized graph representations. For example, as shown in Figure~\ref{fig:graph_examples}(A), an image is parsed using OpenCV~\cite{opencv}, where each pixel’s RGB vector becomes a vertex. Edges are added based on four-way pixel adjacency, with weights computed from Euclidean distances between pixel values. After mutation, \tool reconstructs the image from the modified graph for execution.

{\em 3. Graph Mutator.}
\tool applies 20 mutation operators to graph elements, guided by neighbor similarity to preserve semantic structure. For example, in a point cloud representing a car, each vertex encodes a 3D point and edges reflect nearest-neighbor relationships. Vertices on adjacent surfaces, such as the hood and windshield, are more likely to receive similar perturbations, preserving local geometry. \tool assigns higher mutation probabilities to such neighbors, increasing the likelihood of producing semantically consistent variants.

\begin{figure}[t]
    \centering
    \scalebox{0.9}{
    \includegraphics[width=\linewidth]{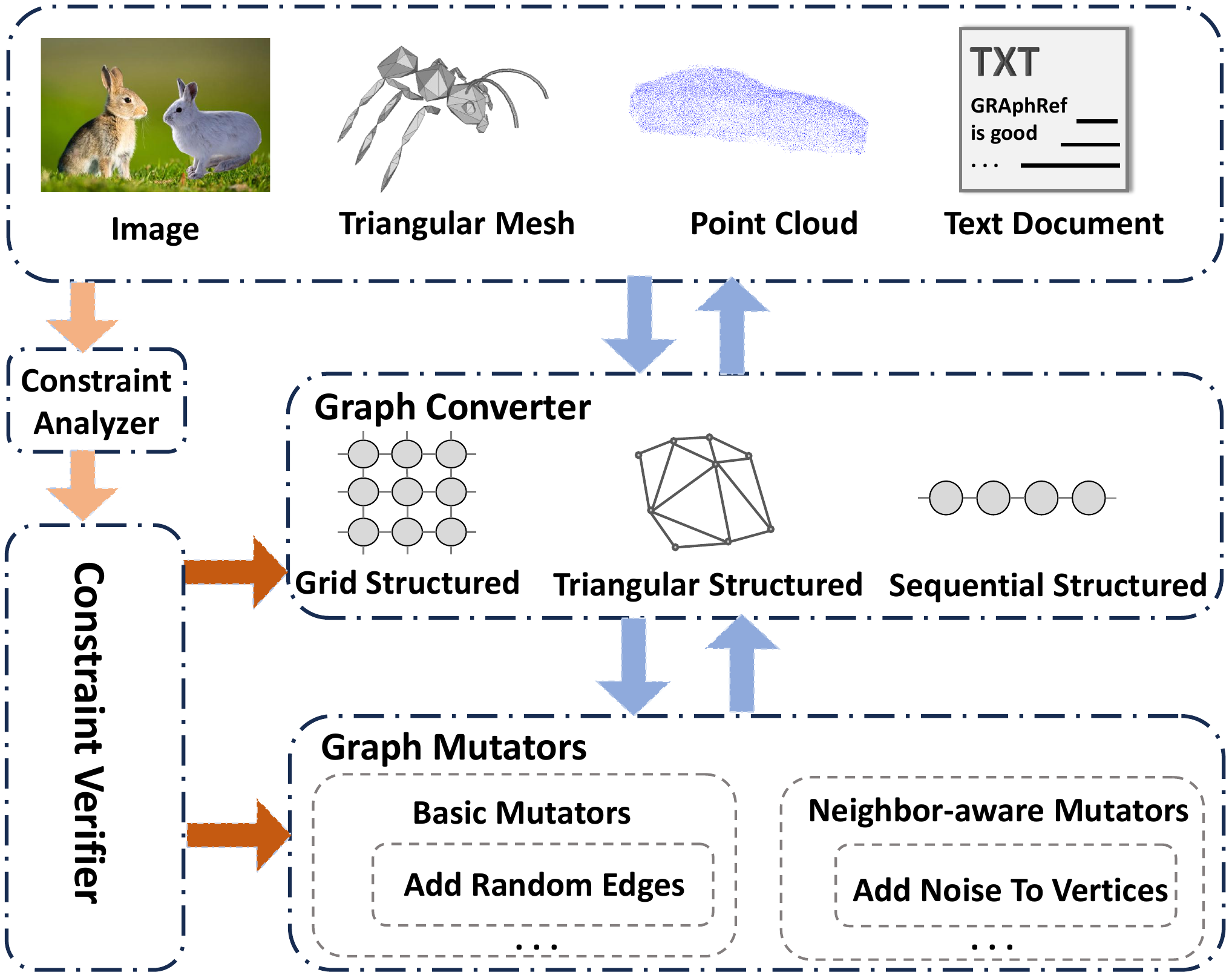}\
    }
    \caption{ \tool Overview;}
    \label{fig:workflow}
\end{figure}

\MyPara{Extensibility.}
In this paper, we target input generation for meshes, a highly structured input type used in various areas based on prior studies~\cite{hanocka2019meshcnn,park2019deepsdf, nasa-vision-2030}.
For example, in computer graphics~\cite{hanocka2019meshcnn,park2019deepsdf}, meshes serve as a fundamental representation, and objects are usually reduced to meshes before rendering; in geographic applications, meshes provide compact representations of terrain data.
Reflecting its broad application and importance, mesh generation is highlighted as a key research direction in NASA's Vision 2030~\cite{nasa-vision-2030}.

Although \tool is engineered for meshes, its underlying technique is general and is applicable to other highly structured inputs, such as images, point clouds, audio, and text, by incorporating appropriate graph representations, as shown in Figure~\ref{fig:graph_examples}.
For example, text files are sequential by nature and thus can be represented as linear graphs where vertices represent words and edges connect adjacent words.
As another example, images and video frames are inherently structured as 2D arrays and thus can be represented by grid graphs.
In other words, \tool can be easily extended to new input types by extending its graph-level APIs, which eliminates the need to reinvent a new input generator.

\MyPara{Execution Plan.}
We plan to evaluate \tool on eight real-world AI systems listed in Table~\ref{tab:mesh_subjects} that process meshes.
In our evaluation, we plan to compare \tool against \ding{202} the classical fuzzing tool, AFL~\cite{afl}; \ding{203} a mesh generator, MeshAttack~\cite{meshattack}; \ding{204} a grammar-based generator, Saffron~\cite{saffron}; \ding{205} two ablated variants of \tool without the constraint-guided refinement, and without neighbor-aware mutation.
We plan to assess its effectiveness in generating valid, structured inputs by measuring the acceptance rate of subject systems without parsing failure, and to use the ablation study to demonstrate the effectiveness of each component. To \ding{206} measure performance and overhead, we will also report the latency breakdown of each major component in \tool, including constraint enforcement, graph conversion, and graph mutation.

\added{\MyPara{Paper Structure.} The remainder of this paper is organized as follows. Section~\ref{sec:background} provides background on highly structured input types to AI systems, such as meshes and point clouds.
Section~\ref{sec:approach} presents the details of the \tool framework we proposed to support this study. Section~\ref{sec:hypotheses} introduces our research questions along with the corresponding hypotheses. Section~\ref{sec:variables} describes the independent, dependent, and confounding variables considered in our study. Section~\ref{sec:execution} outlines our execution plan, including subjects, experimental setup, and evaluation steps. Section~\ref{sec:threats} discusses threats to the validity of our findings. Section~\ref{sec:relatedwork} surveys related work on structured input generation and graph-based mutation. Section~\ref{sec:conclusion} concludes the paper.
}

\section{Background}\label{sec:background}


Emerging applications rely on diverse input types tailored to their functions. Below are common examples:
\begin{itemize}
\item {\bf Meshes and Point Clouds.} Used in 3D vision and graphics for tasks like reconstruction~\cite{park2019deepsdf}, these inputs capture spatial structure.
\item {\bf Images, Audio, and Sensor Data.} Image-based tools in healthcare~\cite{CancerImage}, voice assistants, and autonomous systems like Apollo~\cite{apolloAuto} process sensory inputs to perform perception and control tasks.
\item {\bf Natural Language Text.} Chatbots require diverse language text input to handle conversational queries effectively.
\end{itemize}

Testing these applications is challenging due to the variety and complexity of input. Common approaches include manual input creation, mining real-world data~\cite{SearchInputsForTestingAi}, and building format-specific generators~\cite{Trey2022SemanticImageFuzzing}.

In contrast, \tool provides an extensible framework that models structured inputs as graphs. Its customizable APIs and mutators support a wide range of data types, while constraint-guided refinement ensures both syntactic and semantic validity.

\section{\tool}\label{sec:approach}
\begin{figure}[t]
    \centering
    \scalebox{0.8}{
    \input{sections/DSL}
    }
        \caption{Partial Graph Constraint Grammar}
    \label{fig:DSL}
\end{figure}

\begin{figure}[t]
    \centering
    \scalebox{0.95}{
    \includegraphics[width=\linewidth]{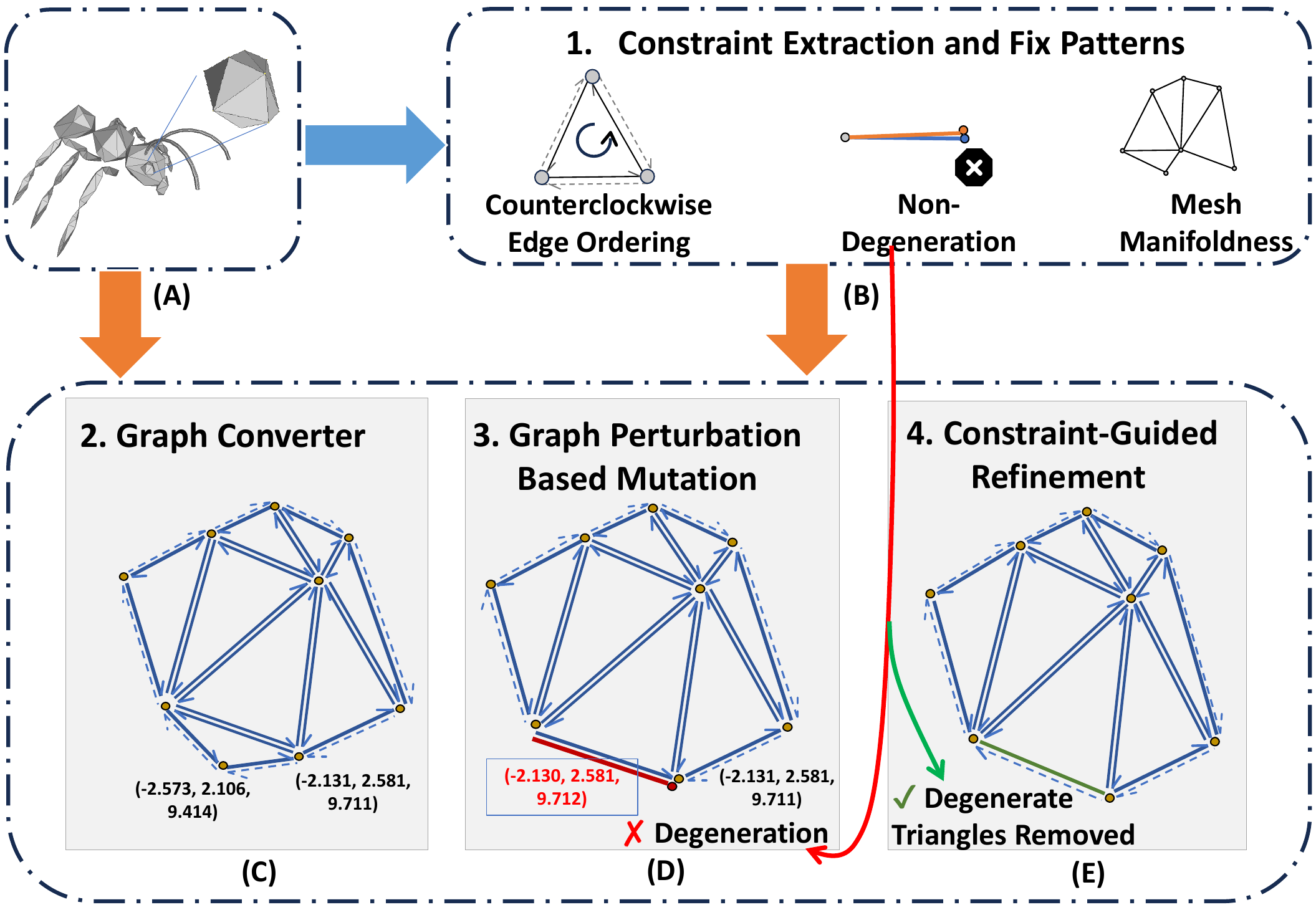}
    }
    \caption{Working Example. (A) Ant-shaped mesh and eye region; (B) Extracted constraints; (C) Constructed graph of eye region; (D) Mutation result with vertex shift; (E) Graph after constraint-guided refinement.}
    \label{fig:workingexample}
    \vspace{-1em}
\end{figure}

To verify our hypotheses, we implemented \tool, a graph-based framework for generating highly structured test inputs with syntactic and semantic guarantees. \tool targets modalities commonly seen in real-world AI systems, such as triangle meshes, point clouds, text, and images. The design and implementation of \tool support our empirical study by enabling controlled mutation and repair of diverse structured inputs.

\tool is based on three core components: (1) graph converter; (2) neighbor-aware mutation engine; and (3) constraint-guided refinement. An overview is shown in Figure~\ref{fig:workflow}.

\MyPara{Graph Converter.}
\tool converts structured data into graphs with semantic attributes. It supports four structural families: sequences (text/audio), grids (images), triangles (meshes), and other relational graphs (\eg point clouds). Each input is preprocessed into a graph format using built-in converters. This abstraction allows the reuse of a unified mutation engine across modalities.

\MyPara{Constraint Encoding.}
\reviewC{To ensure validity over graphs, we define domain-specific structural constraints using an extended version of ISLa~\cite{isla}.} Constraints include properties such as face orientation, area thresholds, vertex connectivity, and manifoldness. These are encoded declaratively and checked before and after mutation. Grammar and an example are shown in Figure~\ref{fig:DSL} and Listing~\ref{lst:dsl_example}.
\begin{minipage}{\linewidth}
\scalebox{0.9}{
\lstdefinelanguage{DSL}{
morekeywords={forall},
sensitive=false,
morecomment=[l]{//},
}

\lstset{
    language=DSL,
  basicstyle=\ttfamily\small,
  keywordstyle=\bfseries,
  captionpos=b,
  mathescape=true,
  numbers=left,
  numberstyle=\sffamily\small\color{gray},
    numbersep=2pt,
  xleftmargin=5em,
}

\begin{lstlisting}[caption={Constraints for Triangle Mesh Graph}, label={lst:dsl_example}]
    G{ // TriangleMeshGraph
      attributes {
        face norm {x, y, z}
      }
      constraints {
        forall (face) {norm.z>0}
        forall (face) {area()>$\epsilon$}
        forall (edge) {
          connected_face()==1 or
          connected_face()==2
        }
        forall (vertex) {
          fan_connected()==true
        }
      }
    }
\end{lstlisting}

}
\vspace{1.5ex}
\end{minipage}

\MyPara{Neighbor-Aware Mutations.}
\tool applies 20 graph mutation operators. To preserve semantic structure, it guides mutation using local neighborhood similarity. For example, in a point cloud of a car, vertices on adjacent surfaces (e.g., hood and windshield) are perturbed similarly to preserve local geometry. This increases the chance of producing valid and semantically coherent inputs.

\begin{table}[t]\small
    \centering    \caption{\reviewB{Constraint Violations on Mesh Graphs Caused by Mutations and Corresponding Refinements (Selected Mutation Types) 
    }}
    \label{tab:constraint_violation_mesh}
    \scalebox{0.88}{
    \begin{tabular}{c|c|c}
        \hline
       \makecell[c]{{\bf Perturbation}} & {\bf Symptoms} & \reviewB{{\bf Fix}} \\
       
        \hline
        
        \makecell[c]{Add Random Noise \\ to Vertices} & \makecell[c]{Triangle vertex \\ position shift}  & \reviewB{\makecell[c]{ Merge duplicate \\ vertices and edges. \\ Remove degenerate triangles.}}\\
        
        \hline 
        
        \makecell[c]{Add Random Noise \\ To Edges} & N/A & \reviewB{N/A} \\
        
        \hline 
        
        \makecell[c]{Randomly Change \\ Vertex Values} & \makecell[c]{Triangle vertex \\ position shift}  & \reviewB{\makecell[c]{ Merge duplicate \\ vertices and edges. \\ Remove degenerate triangles.}} \\
        
        \hline 
        
         \makecell[c]{Randomly Change \\ Edge Values} & N/A & \reviewB{N/A} \\ 
         
         \hline 
         
        \makecell[c]{Randomly Insert \\ Vertices on Edges} & Collinear Vertices & \reviewB{Merge vertices.} \\
        
        \hline 
        
        \makecell[c]{Randomly \\ Add Vertices} & Isolated vertices & \reviewB{\makecell[c]{Edge addition \\ is limited to vertices \\ within the same triangle.}} \\


        
        
        
        
        
       \hline
    \end{tabular}
    }
\end{table}
\MyPara{Constraint-Guided Refinement.} \reviewB{If a mutation causes constraint violations (\eg degenerate triangles), \tool performs targeted repair using predefined strategies. For instance, vertices too close together may be merged, or invalid faces may be removed. This step recovers structural validity and reduces discarded inputs. Table~\ref{tab:constraint_violation_mesh} summarizes supported violations and repairs.}


\section{Research Questions and Hypotheses}\label{sec:hypotheses}
Modern AI systems increasingly rely on highly structured inputs, such as 3D meshes, point clouds, and images, to perform tasks in domains like robotics, medical diagnostics, and autonomous vehicles. These structured inputs exhibit rich topological, geometric, or syntactic properties, which make their generation particularly challenging for existing fuzzing or input generators.

In this study, we propose \tool, a graph-based framework designed for mutating and generating highly structured inputs through a unified graph representation. It combines a neighbor-aware mutation strategy with a constraint-guided refinement mechanism to preserve semantic integrity and ensure structural validity. 

To guide our investigation, we propose the following research questions and associated hypotheses:

\begin{tcolorbox}[colback=blue!5!white, colframe=blue!75!black, boxrule=0.5pt, top=2pt, bottom=2pt]
\MyPara{RQ1.} To what extent does the graph representation support generalization across structured input formats and ensure the validity of the inputs?
\end{tcolorbox}

\MyPara{Hypothesis H1a} Many highly structured inputs, such as 3D meshes, point clouds, and images, share common graph structures that can be captured using extensible APIs, enabling unification under a single mutation engine.

\MyPara{Rationale.} Despite apparent differences in syntax and format, many structured inputs exhibit topological and relational regularities that can be abstracted as graphs. By leveraging a flexible graph API, \tool aims to support multiple input formats without re-engineering domain-specific generators.

\MyPara{Hypothesis H1b.} Structural constraints, such as connectivity and geometric validity, can be formally encoded and enforced during graph mutation to yield valid inputs.

\MyPara{Rationale.} Highly structured inputs like 3D meshes and point clouds often require conformance to strict domain constraints. We hypothesize that representing these constraints at the graph level and applying constraint-guided refinement enables efficient generation of valid test inputs across multiple domains.

\begin{tcolorbox}[colback=blue!5!white, colframe=blue!75!black, boxrule=0.5pt, top=2pt, bottom=2pt]
\MyPara{RQ2.} How effective is \tool's constraint-guided input refinement?
\end{tcolorbox}

\reviewA{
\MyPara{Hypothesis H2.} Constraint-guided refinement can automatically repair structural violations introduced during mutation while preserving the core intent of the change.}

\reviewA{\MyPara{Rationale.} Mutation strategies may occasionally introduce violations of structural or semantic constraints. Instead of discarding such inputs, \tool attempts to repair them through refinement guided by formalized constraints. We expect this step to recover validity in most cases with minimal deviation from the original mutation.}

\begin{tcolorbox}[colback=blue!5!white, colframe=blue!75!black, boxrule=0.5pt, top=2pt, bottom=2pt]
\MyPara{RQ3.} What is the capability of \tool’s neighbor-aware mutation strategy?
\end{tcolorbox}

\reviewA{\MyPara{Hypothesis H3.} Applying mutations to locally related vertices or edges in the input graph results in semantically coherent input variants, increasing the likelihood of preserving task-relevant behaviors.}

\reviewA{\MyPara{Rationale.} In structured inputs, local graph structure often encodes semantically meaningful relationships. For example, in a point cloud representing a car, each vertex corresponds to a 3D point, and edges encode nearest-neighbor spatial proximity. Vertices on adjacent surfaces, such as the hood and the windshield, are more likely to undergo similar perturbations. By prioritizing structurally similar neighbors during mutation, \tool increases the likelihood of generating semantically consistent inputs, which is especially important for testing AI systems sensitive to geometric coherence.}

\begin{tcolorbox}[colback=blue!5!white, colframe=blue!75!black, boxrule=0.5pt, top=2pt, bottom=2pt]
\MyPara{RQ4.} What is the overhead of each component?
\end{tcolorbox}

\reviewA{\MyPara{Hypothesis H4.} Despite the additional abstraction and constraint enforcement layers, \tool introduces only moderate computational overhead suitable for both batch and interactive testing.}

\reviewA{\MyPara{Rationale.} While \tool performs additional processing, such as graph abstraction, constraint checking, and refinement, we hypothesize that these steps scale linearly with input size and remain efficient in practice. This would make the approach viable for large-scale input generation in testing pipelines.}

\section{Variables}\label{sec:variables}

Our study investigates the effects of \tool's core components, namely its graph-based representation, neighbor-aware mutation, and constraint-guided refinement, on the validity, semantic consistency, and performance of structured input generation. We organize our investigation around four research questions and operationalize them through a set of independent, dependent, and confounding variables. 

\MyPara{Independent Variables.}
We define three independent variables that correspond to the core configurable mechanisms in \tool:

\begin{itemize}
    \item \textbf{Constraint-Guided Refinement (CGR):} A binary variable (enabled / disabled) indicating whether \tool performs refinement after mutation to enforce domain-specific constraints.
    
    \item \textbf{Neighbor-Aware Mutation (NAM):} A binary variable (enabled / disabled) controlling whether mutation probabilities are biased toward locally adjacent vertices in the input graph.
    
    \item \textbf{Input Type (IT):} A nominal variable representing the domain of the structured input (e.g., 3D mesh, point cloud), each encoded via \tool's extensible graph representation.
\end{itemize}

\MyPara{Dependent Variables.}
The outcomes of interest in our evaluation are captured via the following dependent variables:

\begin{itemize}
    \item \textbf{Valid Input Rate (VIR):} A ratio-scale variable measuring the proportion of generated inputs that satisfy all domain-specific validity checks.
    
    \item \textbf{Semantic Preservation Score (SPS):} \reviewB{ An interval-scale variable that quantifies the degree to which input semantics are preserved after mutation. It is computed as the percentage of valid mutated inputs for which the downstream model produces the same top-1 prediction label as for the original input. Higher values indicate stronger preservation of semantic structure under mutation.}
    
    \item \textbf{Mutation Diversity (MD):} A ratio-scale variable capturing the structural diversity of generated inputs, computed using graph edit distance or spatial dissimilarity metrics.
    
    \item \textbf{Execution Time per Input (ETI):} A ratio-scale variable representing the average wall-clock time (in milliseconds) required to generate and validate input batches.
\end{itemize}

\MyPara{Confounding Variables.}
To ensure valid causal inference, we identify and control for the following confounding variables:

\begin{itemize}
    
    \item \textbf{Mutation Budget (MB):} A ratio-scale variable defining the number of mutation operations applied per input. This is fixed across conditions to isolate the effects of CGR and NAM.
\end{itemize}

These variables provide a structured framework for evaluating how \tool’s design decisions affect input validity, semantic quality, and runtime efficiency, while controlling for external factors that may influence outcomes.

\section{Execution Plan}\label{sec:execution}

\begin{table}[t]
\centering
\caption{\reviewA{AI systems under tests used as evaluation subjects. All models take triangle meshes as input.}}
\label{tab:mesh_subjects}
\scalebox{1}{
\begin{tabular}{cll}
\toprule
\textbf{Model} & \textbf{Task} & \reviewB{\textbf{Dataset}} \\
\midrule
\reviewA{MeshCNN}~\cite{hanocka2019meshcnn} & \makecell[l]{Mesh classification \\ and segmentation} & \reviewB{\makecell[l]{FAUST~\cite{faust}, \\ ShapeNet~\cite{shapenet}}} \\ \hline
\reviewA{HodgeNet}~\cite{hodgenet} & Spectral classification & \reviewB{ModelNet40~\cite{modelnet40}} \\ \hline
\reviewA{MeshSDF}~\cite{meshSDF} & \makecell[l]{Signed distance \\ function prediction} & \reviewB{ShapeNet~\cite{shapenet}} \\ \hline
\reviewA{Point2Mesh}~\cite{point2mesh} & Surface reconstruction & \reviewB{COSEG~\cite{coseg1}} \\ \hline
\reviewA{MeshWalker}~\cite{meshwalker} & Mesh sequence learning & \reviewB{SMAL~\cite{smal}} \\ \hline
\reviewA{DeepGCNs}~\cite{deepgcn} & \makecell[l]{Mesh-based \\ graph convolution} & \reviewB{ModelNet40~\cite{modelnet40}} \\ \hline
\reviewA{GEM}~\cite{gem} & \makecell[l]{Geometry encoding \\ and segmentation} & \reviewB{Human3.6M~\cite{human36m}} \\ \hline
\reviewA{SpiderCNN}~\cite{spidercnn} & Mesh classification & \reviewB{ModelNet10~\cite{modelnet40}} \\
\bottomrule
\end{tabular}
}
\end{table}

We evaluate \tool using eight real-world AI systems that operate on structured 3D mesh inputs. These systems are drawn from widely used academic benchmarks and represent various downstream AI tasks, including classification, segmentation, and reconstruction. All selected models take triangle meshes as input and apply graph- or geometry-aware reasoning in their pipelines. We include systems that differ in model architecture and dataset source to ensure diversity in test conditions. \reviewA{Table~\ref{tab:mesh_subjects} lists the subjects used in our evaluation.}

\reviewB{All systems are pre-trained on their respective datasets. We select a fixed number of representative mesh samples from each dataset to serve as seed inputs for input generation and testing. These seeds span multiple object categories and levels of geometric complexity.}

We design a controlled experiment to test the hypotheses and answer the four research questions described in Section~\ref{sec:hypotheses}. All experiments are conducted on a machine running Ubuntu 24.04 with 32 GB RAM and an Intel i9-13900HX processor.

\subsection{Step 1: Seed Selection and Preprocessing}

We select a total of 96 triangle mesh inputs, sampled uniformly across 12 object categories in the ShapeNetCore dataset~\cite{shapenet}. These meshes serve as the initial seed inputs for all test generation tools. 

\subsection{Step 2: Baseline and Tool Configuration}

We compare \tool against five baselines:
\begin{itemize}
    \item \textsc{AFL}~\cite{afl}: A generic mutation-based fuzzer that operates on raw bytes.
    \item \textsc{GraphGen}: An ablated version of \tool without constraint-guided refinement.
    \item \textsc{GraphNoNeighbor}: A version of \tool that disables neighbor-aware mutation.
    \item Saffron~\cite{saffron}: A grammar-based generator requiring a handcrafted grammar.
    \item MeshAttack~\cite{meshattack}: A mesh-specific adversarial perturbation approach.
\end{itemize}

All tools are run for 1800 seconds per seed input. For fairness, mutation budgets and thread counts are standardized across tools.

\subsection{Validity Evaluation (RQ1)}

We run each tool against the selected subjects and count the number of generated inputs that are accepted by the AI systems without triggering parsing errors. Acceptance is determined based on runtime exceptions, file load failures, or invalid topology flags.

\subsection{Constraint-Guided Refinement Evaluation (RQ2)}

We evaluate the effectiveness of \tool's constraint-guided refinement by disabling the verifier module (\textsc{GraphGen}) and measuring the difference in validity rate and constraint violation types. We perform this ablation for each subject system using the same set of seed meshes and runtime limits.

\subsection{Neighbor-Aware Mutation Evaluation (RQ3)}

To assess semantic preservation, we compare the classification outputs of subject systems on original and mutated meshes with and without neighbor-aware mutation (\textsc{GraphNoNeighbor}). For each valid mutated input, we check whether the model’s prediction matches the prediction on the original input. The percentage of matching predictions is reported as the semantic consistency rate.

\subsection{Performance Overhead Analysis (RQ4)}

We instrument \tool to record the time spent in each major stage: graph construction, neighbor selection and mutation, and constraint refinement. We report the average per-input execution time and compare it to the baseline generators to evaluate runtime feasibility.


\section{Threats to Validity}\label{sec:threats}
\reviewA{

\noindent{\bf Subjects and Data Types.} \reviewC{The experiments were conducted on a specific set of subjects and data types. While the absolute values of input validity, execution time, and symptoms may vary with unseen subjects, it is anticipated that the advantages of \tool in terms of acceleration and semantic preservation will hold when applied to other platforms.}

\noindent{\bf Time Limit.} We empirically set 30 minutes as the time limit for fuzzing. Longer execution time may expose more symptoms or more execution paths as suggested in~\cite{evaluatefuzzing}; however, this time limit is reasonable because we did not see any increase in new types of errors with a higher time limit for S1-S5.

\noindent{\bf Utilized Structures.} The evaluation was conducted using specific graph structures defined within the tool’s framework. Various graph representations can exist for the same data type. For example, images can be modeled at the pixel level for fine-grained details or at the object level for higher-level features. These differing representations can influence the outcomes of \tool, as the absolute values of results may vary with the chosen graph structure. Currently, \tool requires a manually pre-selected graph structure model to reflect the data's inherent structure.

}

\section{Related Work}\label{sec:relatedwork}

\MyPara{Fuzz Testing.}
Fuzz testing explores program behavior by generating diverse inputs, but bit-level mutations often fail to produce valid structured inputs. To address this, prior work proposes token-level~\cite{274551-token-level-fuzzing-security21} and probabilistic mutations~\cite{probabilityFuzzing}. In contrast, \tool models structured inputs as graphs, applies graph perturbations, and preserves structures and semantics through constraint validation on graphs.

Grammar-based fuzzers improve efficiency by reducing invalid inputs. Saffron~\cite{saffron} adapts user-provided grammars based on learned input acceptance. Other works~\cite{superion} use inferred or grammar-specific mutations for structured domains. ISLa~\cite{isla} extends context-free grammars with semantic constraints to generate inputs that are both syntactically and semantically valid. In contrast, the tool unifies input formats as graphs and extends ISLa to graphs, utilizing a reusable mutation engine. 

\MyPara{Graph-based Mutation.}
Graph representations have been used in domains such as dataflow analysis~\cite{Harrison2022GraphFuzz}. 
GraphFuzz models function traces as graphs and mutates execution paths. However, prior tools are domain-specific and do not generalize to multimodal inputs. \tool generalizes graph-based mutation by leveraging intrinsic structure in diverse input types for automated, reusable fuzzing.

\MyPara{Input Constraints and Validity.}
Constraint-guided input generation ensures syntactic and semantic correctness. Tools like NNSmith~\cite{Jiawei2023NNSmith} and NeuRI~\cite{Jiawei2023NeuRI} enforce constraints on neural networks and tensor shapes. \tool encodes input validity as graph-level constraints and checks them during mutation to ensure generated inputs remain valid.

\MyPara{AI-based Input Generation.}
Learning-based methods~\cite{adversarialADS, llminputgen} use neural models to generate structured inputs, offering flexibility beyond static grammars. However, they often lack explicit constraint handling, require costly training and inference, and still depend on predefined templates or constraints. As a result, they have limited capability in generating structurally or semantically diverse inputs. \tool offers a lightweight alternative by combining symbolic constraints with structure-aware mutation.

\section{Conclusion}\label{sec:conclusion}

\tool is a framework developed to evaluate our hypotheses on structured input generation. It models inputs as graphs to guide semantically meaningful mutations and applies constraint-based refinement to preserve input validity. The forthcoming evaluation will assess its effectiveness across diverse AI systems.

\section{Data Availability}

We provide access to our anonymized artifacts at \url{https://doi.org/10.5281/zenodo.15532130}.

\balance

\newpage
\bibliographystyle{IEEEtran}
\bibliography{reference,issta}

\end{document}